\documentclass[12pt]{article}
\usepackage{graphicx}
\usepackage{amssymb}
\usepackage{cite}
\usepackage{bm}
\begin{document}
\setlength{\unitlength}{1mm}
\textwidth 15.0 true cm 
\headheight 0 cm
\headsep 0 cm 
\topmargin 0.4 true in
\oddsidemargin 0.25 true in
\input epsf

\newcommand{\beq}{\begin{equation}}
\newcommand{\eeq}{\end{equation}}
\newcommand{\be}{\begin{eqnarray}}
\newcommand{\ee}{\end{eqnarray}}
\renewcommand{\vec}[1]{{\bf #1}}
\newcommand{\vecg}[1]{\mbox{\boldmath $#1$}}
\newcommand{\grpicture}[1]
{
    \begin{center}
        \epsfxsize=200pt
        \epsfysize=0pt
        \vspace{-5mm}
        \parbox{\epsfxsize}{\epsffile{#1.eps}}
        \vspace{5mm}
    \end{center}
}

\begin{flushright}

\end{flushright}

\vspace{0.5cm}

\begin{center}

{\Large\bf   Ghost-free higher-derivative theory}

\vspace{1cm}

{\Large A.V. Smilga} \\

\vspace{0.5cm}

{\it SUBATECH, Universit\'e de
Nantes,  4 rue Alfred Kastler, BP 20722, Nantes  44307, France. }
\footnote{On leave of absence from ITEP, Moscow, Russia.}\\

\end{center}

\bigskip

\begin{abstract}
We present an example of the quantum system with higher derivatives  in the Lagrangian, which is ghost-free:
the spectrum of the Hamiltonian is bounded from below and unitarity is preserved. 
\end{abstract}

\section{Introduction.}
Since the classical paper of Pais and Uhlenbeck \cite{PU} it was generally believed that field
theories with higher derivatives (HD) in the Lagrangian are intrinsically sick --- the spectrum of such a theory
necessarily involves ghost states, which break unitarity and/or causality.

To understand the nature of this problem, one need not to study field theories. It is clearly seen in toy models with finite
number of degrees of freedom. Consider e.g. the Lagrangian
 \be
\label{Om4}
{ L} \ =\ \frac 12 \ddot q^2 - \frac {\Omega^4}2  q^2  \ .
 \ee 
It is straightforward to see that four independent solutions of the corresponding classical equations of motion are 
$q_{1,2}(t) = e^{\pm i\Omega t}$, $q_3(t) = e^{-\Omega t}$ and $q_4(t) = e^{\Omega t}$. The exponentially rising solution
$q_4(t)$ displays instability of the classical vacuum $q=0$. The quantum Hamiltonian of such a system is not hermitian
and the evolution operator $e^{-i\hat H t}$ is not unitary.

 This vacuum instability is characteristic for all {\it massive} HD field theories  --- the 
dispersive equation has complex solutions in this case for small enough momenta. 
But for intrinsically massless (conformal) field theories
the situation is different. Consider the Lagrangian
   \be
\label{L4mix}
{     L} \ =\ \frac 12 (\ddot q  + \Omega^2 q )^2 -  \frac \alpha 4 q^4 - \frac {\beta}2 q^2 \dot q^2 \ .
 \ee
Its quadratic part can be obtained from the HD field theory Lagrangian ${ L} = (1/2) \phi \Box^2 \phi$
involving massless scalar field, when restricting it on the modes with a definite momentum $\vec{k}\ \  
(\Omega^2 = \vec{k}^2)$.
 If neglecting the nonlinear terms in (\ref{L4mix}), the solutions of the classical equations of motion
$q(t) \sim e^{\pm i\Omega t}$ and $q(t) \sim te^{\pm i\Omega t}$ do not involve exponential instability, but include only
comparatively ``benign'' oscillatory solutions with linearly rising  amplitude. 

We showed in \cite{duhi} that, when nonlinear terms in Eq.(\ref{L4mix}) are included, an island of stability in the 
neighbourhood of the classical vacuum
\footnote{Usually, the term classical vacuum is reserved for the point in the configuration (or phase) space with 
minimal energy. For HD theories and in particular for the theory (\ref{L4mix}) the classical energy functional is not bounded
from below and by ``classical vacuum'' we mean simply a stationary  solution to the classical equations of motion.} 
 \be
 \label{vac}
q = \dot q = \ddot q = q^{(3)} = 0
 \ee
exists in a certain range of the parameters $\alpha, \beta$. In other words, when initial conditions are chosen at the vicinity
of this point, the classical trajectories $q(t)$ do not grow, but display a decent oscillatory behaviour. 
This island  is surrounded by the sea of instability, however. For generic initial conditions, the trajectories become singular
--- $q(t)$ and its derivatives reach infinity in a finite time.

Such a singular behaviour of classical trajectories often means trouble also in the quantum case. A well-known example 
when it does is the problem of $3D$ motion in the potential
 \be
\label{potr2}
V(r) \ =\ - \frac \gamma {r^2}\ .
  \ee
The classical trajectories where the particle falls to the centre (reach the singularity $r=0$ in a finite time) are abundant.
This occurs when $L > \sqrt{2m\gamma}$, where $L$ is the classical angular momentum.
And it is also well known that, if $m\gamma > 1/4$, the quantum problem is not very well defined: the eigenstates with arbitrary 
negative energies exist and the Hamiltonian does not have a ground state. 

 The bottomlessness of the Hamiltonian is not, however, a necessary consequence of the fact that the classical problem
involves singular trajectories. In the problem (\ref{potr2}), the latter are present for all positive $\gamma$, 
but the quantum ground state disappears only when $\gamma$ exceeds the boundary value $1/(4m)$.  

 The main statement of this paper is that the system (\ref{L4mix}) exhibits a similar behaviour. If both $\alpha$ and $\beta$
are nonnegative (and at least one of them is nonzero), the quantum Hamiltonian has a bottom and the quantum problem
is perfectly well defined even though some classical trajectories are singular.

\section{Free theory}

Consider the Lagrangian $(\ddot q + \Omega^2 q)^2/2$ obtained from (\ref{L4mix}) by suppressing  the interaction terms. 
The quantum dynamics of such a theory was studied in \cite{DM}. It is instructive to consider first the Lagrangian
  \be
 \label{om12}
 { L} \ =\ \frac 12 \left[  \ddot q^2 - (\Omega_1^2 + \Omega_2^2) \dot q^2 + \Omega_1^2 \Omega_2^2 q^2 \right]
  \ee
and look what happens in the limit $\Omega_1 \to \Omega_2$. The main observation of Ref.\cite{DM} was that this limit
is singular. When $\Omega_1 > \Omega_2$, the spectrum of the theory (\ref{om12}) is 
\be
\label{spec12}
E_{nm} = \left(n+ \frac 12 \right) \Omega_1 - \left( m + \frac 12 \right) \Omega_2 
\ee
with nonnegative integer $n,m$. On the other hand, when $\Omega_1 = \Omega_2 = \Omega$, the spectrum is
 \be
 \label{spec}
E_n =n\Omega
\ee
with generic integer $n$. In both cases, the quantum Hamiltonian has no ground state, but in the limit of equal
frequencies the number of degrees of freedom is reduced in a remarkable way: instead of two quantum numbers $n,m$
(The presence of two quantum numbers is natural --- the phase space of the system (\ref{om12}) is 4--dimensional
giving two pairs $(p_{1,2}, q_{1,2})$ of canonic variables.), we are left with only one quantum number $n$.

This deficiency of the number of eigenstates compared to natural expectations would not surprise a mathematician. A generic
$2\times 2$ matrix has two different eigenvectors. But the Jordan cell 
$\left( \begin{array}{cc} 1 & 1 \\ 0 & 1 \end{array}
\right) $
has only {\it one} eigenvector $\propto \left( \begin{array}c 1 \\ 0 \end{array} \right) $. 
The statement is therefore that in the 
limit $\Omega_1 = \Omega_2$ our Hamiltonian represents a kind of generalized Jordan cell.
In what follows we partially repeat the analysis of \cite{DM}, but make it much more explicit. This will allow
us to understand this ``Jordanization'' phenomenon in very clear terms and  prepare us to the analysis of the
interacting theory (\ref{L4mix}).

To begin with, let us construct the canonical Hamiltonian corresponding to the Lagrangian (\ref{om12}). This can be done 
using the general Ostrogradsky formalism \cite{Ostr}
\footnote{See e.g. \cite{Ham} for its detailed pedagogical description.}. 
For a Lagrangian like (\ref{om12}) involving $q, \dot q$, and $\ddot q$, it consists in introducing the new variable
$x = \dot q$ and writing the Hamiltonian $H(q,x; p_q, p_x)$ in such a way that the classical Hamilton equations
of motion would coincide after excluding the variables $x, p_x, p_q$ with the equations of motion
 \be
\label{eqmot}
q^{(4)} + (\Omega_1^2 + \Omega_2^2) \ddot q + \Omega_1^2 \Omega_2^2 q \ =\ 0
 \ee
derived from the Lagrangian (\ref{om12}). This Hamiltonian has the following form
  \be
\label{Ham12}
H \ =\ p_q x + \frac {p_x^2}2 + \frac {(\Omega_1^2 + \Omega_2^2) x^2}2 - \frac {\Omega_1^2 \Omega_2^2 q^2}2 \ .
  \ee
For example, the equation $\partial H/\partial p_q = \dot q$ gives the constraint $x = \dot q$ , etc. 

Assume for definiteness $\Omega_1 > \Omega_2$. 
The Hamiltonian (\ref{Ham12}) can then be brought into diagonal form by the following canonical transformation
   \be
 \label{canon}
q &=& \frac {a_1}{\sqrt{2\Omega_1 (\Omega_1^2 - \Omega_2^2)}} \ +\  \frac {a_2}{\sqrt{2\Omega_2 (\Omega_1^2 - \Omega_2^2)}}
\ +\ {\rm h.c.}\ , \nonumber \\
x &=& - \frac {i\Omega_1 a_1}{\sqrt{2\Omega_1 (\Omega_1^2 - \Omega_2^2)}} \ +\  
\frac {i\Omega_2 a_2}{\sqrt{2\Omega_2 (\Omega_1^2 - \Omega_2^2)}}
\ +\ {\rm h.c.} \ ,\nonumber \\
p_x &=& -\frac {\Omega_1^2 a_1}{\sqrt{2\Omega_1 (\Omega_1^2 - \Omega_2^2)}} \ -\  
\frac {\Omega_2^2 a_2}{\sqrt{2\Omega_2 (\Omega_1^2 - \Omega_2^2)}}
\ +\ {\rm h.c.} \ , \nonumber \\
p_q &=& \frac {i \Omega_1 \Omega_2^2 a_1}{\sqrt{2\Omega_1 (\Omega_1^2 - \Omega_2^2)}} \ -\  
\frac {i \Omega_2 \Omega_1^2 a_2}{\sqrt{2\Omega_2 (\Omega_1^2 - \Omega_2^2)}}
\ +\ {\rm h.c.} \ .
 \ee
We obtain 
  \be
\label{Hdiag}
H \ =\ \Omega_1 a_1^* a_1 - \Omega_2 a_2^* a_2 \ .
  \ee
The classical dinamics of the Hamiltonian (\ref{Hdiag}) is simply $a_1 \propto e^{-i\Omega_1 t}$, 
$a_2 \propto e^{i\Omega_2 t}$. Quantization of this Hamiltonian gives the spectrum (\ref{spec12}).
The negative sign of the second term in (\ref{Hdiag}) implies the negative sign of the corresponding kinetic term,
which is usually interpreted as the presence of the ghost states (the states with negative norm) in the spectrum.
We prefer to keep the norm positive definite, with the creation and annihilation operators $a_{1,2}, a_{1,2}^\dagger$
(that correspond to the classical variables   $a_{1,2}, a_{1,2}^*$) satisfying the usual commutation relations
  \be
 \label{commut}
[a_1, a_1^\dagger] \ =\ [a_2, a_2^\dagger]\ =\ 1\ .
 \ee
However, irrespectively of whether the metric is kept positive definite or not and the world ``ghost'' is used or not, the
spectrum (\ref{spec12}) does not have a ground state and, though the spectral problem for the free Hamiltonian
(\ref{Hdiag}) is perfectly well defined, the absence of the ground state leads to a trouble, the falling to the centre
phenomenon when switching on the interactions. 
\footnote{A characteristic feature of this phenomenon is that classically trajectories reach singularity in a finite
time while the quantum spectrum  involves a {\it continuum} of states with arbitrary low energies
\cite{fc}.
In our case, the ``centre'' is not a particular point in the configuration (phase) space, but rather
its boundary at infinity, but the physics is basically the same.}

We are interested, however, not in the system (\ref{om12}) as such, but rather in this system in the limit
 $\Omega_1 =\Omega_2$.
The latter is not so trivial because we are {\it not} allowed to simply set $\Omega_1 =\Omega_2$ in (\ref{Hdiag}) --- the
canonical transformation (\ref{canon}) is singular at this point. The best way to see what happens in this limit
is the explicit one. We will write down the expressions for the wave functions of the states (\ref{spec12}) and
explore their behaviour in the equal frequency limit. 

This can be done by starting from the oscillator wave functions in the basis (\ref{Hdiag}) and performing the canonical
transformation (\ref{canon}). An alternative, the most direct way is substitute the operators $-i \partial/\partial x,
\ -i \partial /\partial q$ for $p_x$ and $p_q$ in Eq. (\ref{Ham12}) and to search for the solutions of the Schr\"odinger
equation in the form
  \be
 \label{AnsPsi}
\Psi(q,x)\ =\ e^{-i\Omega_1 \Omega_2 qx} \exp\left\{ - \frac \Delta 2 \left( x^2 + 
\Omega_1 \Omega_2 q^2 \right) \right\} \phi(q,x)\ ,
 \ee
where $\Delta = \Omega_1 - \Omega_2$. Then the operator acting on $\phi(q,x)$ is
 \be
\label{Opphi}
 \tilde H \ =\ - \frac 12 \frac {\partial^2}{\partial x^2} + \left( \Delta x + i\Omega_1 \Omega_2 q \right) 
\frac \partial {\partial x} - ix \frac \partial {\partial q} + \frac \Delta 2\ .
 \ee
It is convenient to introduce 
 \be
z = \Omega_1 q + ix\ ,\ \ \ \ \  \ \ 
u = \Omega_2 q - ix \ ,
 \ee
after which the operator (\ref{Opphi}) acquires the form
  \be
 \label{Hhol}
 \tilde H(z,u) \ =\ \frac 12 \left( \frac \partial {\partial z} - \frac \partial {\partial u} \right)^2 + 
\Omega_1 u \frac \partial {\partial u} - \Omega_2 z \frac \partial {\partial z} + \frac \Delta 2 \ .
  \ee 
The holomorphicity of $ \tilde H(z,u)$ means that its eigenstates are holomorphic functions $\phi(z,u)$. An obvious
eigenfunction with the eigenvalue $\Delta/2$  is $\phi(z,u) = $ const. Further, if assuming  $\phi$ to be the function of
only one holomorphic variable $u$ or $z$, the equation $\tilde H \phi = E \phi$ acquires the same form as for the equation
for the preexponential factor in the standard oscillator problem. Its solutions are Hermit polynomials,
 \be
\label{n0i0m}
\phi_n(u) &=& H_n(i\sqrt{\Omega_1} u ) \equiv H_n^+ ,\ \ \ \ \ \ E_n = \frac \Delta 2 + n\Omega_1 \ , \nonumber \\
\phi_m(z) &=& H_m(\sqrt{\Omega_2} z ) \equiv H_m^- ,\ \ \ \ \ \ E_m = \frac \Delta 2 - m \Omega_2 \ .   
  \ee
 
The solutions (\ref{n0i0m}) correspond to excitations of only one of the oscillators while another one is in its 
ground state.
For sure, there are also the states where both oscillators are excited. One can be directly convinced that the functions
  \be
\label{sumHerm}
\phi_{nm}(u, z) &=& \sum_{k=0}^m \left( \frac {i\Delta }{4 \sqrt{\Omega_1 \Omega_2}} \right)^k 
\frac {(n-m+k+1)!}{(m-k)! k!} H^+_{n-m+k} H^-_k,\ \ \ \  m\leq n \ , \nonumber \\
\phi_{nm}(u, z) &=& \sum_{k=0}^n \left( \frac {i\Delta }{4 \sqrt{\Omega_1 \Omega_2}} \right)^k 
\frac {(m-n+k+1)!}{(n-k)! k!} H^+_{k} H^-_{m-n+k},\ \ \ \  m >  n  \nonumber \\
&\ &
 \ee
are the eigenfunctions of the operator (\ref{Hhol}) with the eigenvalues (\ref{spec12}). Multiplying the polynomials
(\ref{sumHerm}) by the exponential factors as distated by Eq.(\ref{AnsPsi}), we arrive at the normalizable
wave functions of the Hamiltonian (\ref{Ham12}).

We are ready now to see what happens in the limit $\Omega_1 \to \Omega_2$ ($\Delta \to 0$). Two important observations
are in order. 

\begin{itemize}

\item The second exponential factor in (\ref{AnsPsi}) disappears and the wave functions cease to be normalizable. This
is unexpected and rather unusual. Indeed, we do not know any other physical quantum system with discrete
spectrum where wave function would not be normalizable. On the other hand, it is not so surprising from mathematical 
viewpoint.
Indeed, the operator (\ref{Hhol}) has discrete spectrum and polynomial, not normalizable eigenfunctions even if the 
frequencies
are not equal
\footnote{The boundary condition leading to discretization of the spectrum is in this case not the requirement of 
normalizability
of the wave function, but the requirement that it grows not faster than a polynomial.}
The same concerns the familiar one-dimentional Hermit operator $(1/2) \partial^2 /\partial z^2 - 
\Omega \, z \partial /\partial z$
and many its Fuchs relatives. Actually, nothing forbids the original Hamiltonian (\ref{Ham12}) to have the same nature and, 
in the equal frequencies limit, it has it.

\item We see that in the limit $\Delta \to 0$, only the first terms survive in the sums (\ref{sumHerm}) and we obtain
  \be
 \label{limHerm}
 \lim_{\Delta \to 0} \phi_{nm} & \sim & H^+_{n-m} \ \ \ \ \ \ \ \ , \ \ \ \ \ \ m \leq n \nonumber \\
\lim_{\Delta \to 0} \phi_{nm}  & \sim & H^-_{m-n} \ \ \ \ \ \ \ \ , \ \ \ \ \ \ m > n\ . 
  \ee
 In other words, the wave functions depend only on the difference $n-m$, which is {\it the only} relevant quantum number
in the limit $\Omega_1 = \Omega_2$.

\end{itemize}

 As this phenomenon is rather unusual and very important for us, let us spend few more words to clarify it. 
Suppose $\Omega_1$ is very close to $\Omega_2$, but still not equal. Then the spectrum includes the sets of nearly 
degenerate states. For example, the states $\Psi_{00}, \Psi_{11}, \Psi_{22}$, etc have the energies $\Delta/2, 3\Delta/2,
5\Delta/2$, etc, which are very close. In the limit $\Delta \to 0$, the energy of all these states coincides, but rather
than having an infinite number of degenerate states, we have only one state: the wave functions 
 $\Psi_{00}, \Psi_{11}, \Psi_{22}$, etc simply {\it coincide} in this limit  by the same token as the eigenvectors
of the matrix  $\left( \begin{array}{cc} 1 & 1 \\ \Delta & 1 \end{array}
\right) $ coincide in the limit $\Delta \to 0$. 

The fact that some matrices have a reduced number of eigenvectors compared to a generic case is very well known. 
One of the reasons why it was never considered relevant for physical spectral problems is that the Jordan matrix
is not symmetric and seems not to be a viable model for a physical Hermitian Hamiltonian. However, it was recently
understood \cite{Bender} that {\it any} matrix (and any Hamiltonian) with real eigenvalues can be rendered
Hermitian if defining the norm in an appropriate way. Indeed, the Hamiltonian (\ref{Ham12}) in the equal frequencies
limit can be represented as a Jordan matrix in the basis with a special non-diagonal metric \cite{DM}.

\section{Interacting theory.}

When $\Omega_1 = \Omega_2$, $u = \bar z$ and the operator (\ref{Hhol}) acquires the form
 \be
\label{Hzzbar}
 \tilde H(z, \bar z) \ =\ \frac 12 \left( \frac \partial {\partial \bar z} -  \frac \partial {\partial  z} \right)^2 +
\Omega \left( z \frac \partial {\partial \bar z} -  \bar z \frac \partial {\partial  z} \right)\ .
 \ee
Its eigenfunctions are either holomorphic or antiholomorphic Hermit polynomials. Let us deform (\ref{Hzzbar}) by
adding there the quartic term $\alpha z^2 \bar z^2$ with positive $\alpha$. Note first of all that it cannot be treated
as a perturbation, however small $\alpha$ is: the wave functions are not normalizable and the matrix elements
of  $\alpha z^2 \bar z^2$ diverge. But one can use the variational approach. Let us take the Ansatz 
  \be
 \label{Ansvar}
 |{\rm var} \rangle \ =\ z^n e^{-Az \bar z} \ ,
 \ee
where $A,n$ are the variational parameters. The matrix element of the unperturbed quadratic Hamiltonian
(\ref{Hzzbar}) over the state (\ref{Ansvar}) is
  \be
\label{var2}
\langle {\rm var} | \tilde H | {\rm var} \rangle \ =\ \frac {A(n+1)}2 - \Omega n\ . 
  \ee 
Obviously, by choosing $n$ large enough and $A$ small enough, one can make it as close to $-\infty$ as
one wishes. The bottom is absent and one cannot reach it. For the deformed Hamiltonian, the situation is different, however.
We have
 \be
 \label{Evar}
E^{\rm var}(n.A) = \langle {\rm var} | \tilde H + \alpha z^2 \bar z^2| {\rm var} \rangle \ =\ \frac {A(n+1)}2 - \Omega n
+ \frac {\alpha(n+1)(n+2)}{4A^2}\ . 
  \ee
This function has a global minimum. It is reached when 
  \be
A-\Omega - \frac \alpha{4A^2} \ =\  0 
  \ee
and  $n =  {A^3}/\alpha - 2$.

For small $\alpha \ll \Omega^3$,
 \be
A \approx \Omega,\ \ n \approx \frac {\Omega^3}\alpha, \ \ \ \ {\rm and} \ \ \ \ E^{\rm var} \approx -\frac {\Omega^4}{4\alpha}\ .
 \ee
The smaller is $\alpha$, the lower is the variational estimate for the ground state energy and the ground state energy itself.
In the limit $\alpha \to 0$, the spectrum becomes bottomless. But for a finite $\alpha$, the bottom exists.

Bearing in mind that $z = \Omega q + ix = \Omega q + i \dot q$, the deformation $\alpha z^2 \bar z^2$ amounts to a 
particular combination of the terms $\sim q^4$, $\sim q^2 \dot q^2$, and $\sim \dot q^4$ in the Hamiltonian. For the theory
(\ref{L4mix}) with generic $\alpha, \beta$, the algebra is somewhat more complicated, but the conclusion is the same: in the case 
when the form $\alpha q^4/4 + \beta q^2 x^2/2$ is positive definite, the system has a ground state.

 The requirement of positive definiteness of the deformation is necessary. In the opposite case, choosing the Ansatz 
$$ |{\rm var} \rangle \ \sim  \ (\Omega q + ix)^n \exp\{-Aq^2 - B x^2 \}$$
and playing with $A,B$,   
one can always make the matrix element $\langle$ {\sl var} $|$ {\sl deformation} $|$ {\sl var} $\rangle $ negative, 
which would add to the
negative contribution $-\Omega n$ in the variational energy, rather than compensate it. The bottom is absent in this case.

\begin{figure}[h]
   \begin{center}
 \includegraphics[width=3.0in]{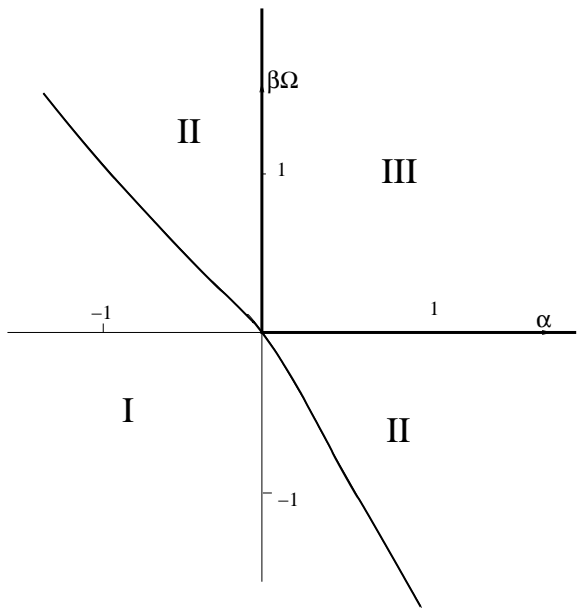}
        \vspace{-2mm}
    \end{center}
\caption{\small Ghost and no-ghost regions. I --- perturbative instability. II --- falling to the centre.
III - Nice Hermitian Hamiltonian endowed with a bottom. }
\label{phase}
\end{figure}

 Combining this   with the findings of Ref.\cite{duhi}, we conclude that the whole plane $(\alpha, \beta )$ is divided 
into 3 regions (see Fig. 1). In the region I, the stationary point $q = \dot q = \ddot q = q^{(3)} = 0$ of  classical phase
space (the would-be classical vacuum) is unstable with respect to small perturbations. In the region II, the island of stability appears
in the classical problem, but some classical trajectories hit the singularity. Speaking of the quantum problem, the Hamiltonian 
does not have a ground state and we are in the falling to the centre situation. In  both cases, unitarity is not preserved.
  Finally, the region III is not distinguished from the
region II as far as the classical problem is concerned (there {\it are} singular trajectories), but the quantum problem is perfectly
well defined: the spectrum has the bottom, the Hamiltonian is Hermitian, and the evolution operator is unitary. 
\footnote{In \cite{duhi}, we dubbed the ghosts appearing in  region I ``malicious'' and the (classical) 
ghosts appearing in regions II and III   ``benign''. We see, however, that the relevant (quantum) 
 watershed passes between the regions I and II on one
hand and the region III on the other hand.} 

\section{Discussion}

What are the implications of the analysis of our toy model (\ref{L4mix}) for field theories ? Is it possible to find a HD field
theory with a well-defined ground state ? We {\it believe} that the answer to this question is positive though cannot 
justify it at the moment. Our faith is based on the fact that we do not see any other way to construct a viable theory 
of quantum gravity.
The argumentation is the following (see \cite{duhi,dragon} for more detailed discussion). 
 \begin{itemize}
 \item  The main conceptional difficulty which has not allowed us up to the present to construct a decent quantum counterpart
to  Einstein's general relativity is its {\it geometric nature}. Time is intertwined there with other coordinates of a curved
manifold and no notion
of universal flat time exists. Standard quantum mechanics (either with finite or with infinite number of degrees of
freedom) {\it is} based on this notion, however. When we write the Schr\"odinger operator and define the evolution operator,
we imply this.\nobreak\footnote{Contrary to what most people think, general relativity has conceptional difficulties even at the
 classical level. The matter
is that Einstein's equations of motion cannot {\it always} be presented in the Cauchy form and, as a result, causality is not always
preserved. There are some exotic solutions involving closed time loops \cite{noncause}. Even though they are unstable 
and are not realized macroscopically, their presence leads to difficulties at the foundational level. These difficulties become 
overwhelming when attempting to construct quantum theory \cite{Isham}. } 
  \item  That is why we suggested \cite{duhi} that the fundamental Theory of Everything is formulated in flat space-time. 
To explain gravity, this bulk should have more than 4 dimensions. Then our curved (3+1)--dimensional Universe may appear as a 
classical solution (3-brane) in the bulk and gravity would have the status of {\it effective} theory on this brane.
 
Our other assumption is that the TOE is not a version of string theory, but rather a conventional field theory. 
It is simply because string
theory does not have today a consistent nonperturbative formulation and one may doubt that such a formulation could
be given. A field theory formulated in more that four dimensions
should involve higher derivatives - otherwise it is not renormalizable. {\it If} one believes that such HD TOE exists,
one should also believe that it is free from ghost-driven paradoxes.

\end{itemize}

What can one say about the form of this theory ? As was argued above, one can only hope to tame the ghosts if the theory is
 {\it conformal}. If not, vacuum is perturbatively unstable at the classical level.  We also want it to be supersymmetric. Otherwise,
the cosmological term --- the energy density of our brane$\equiv$Universe cannot be expected to vanish. The requirement
of superconformal invariance imposes stringent constraints on the theory.
 For example, it  restricts the number of  dimensions of the flat
space-time  where the theory is formulated by $D \leq 6$. Indeed, all superconformal algebras (involving the super-Poincare algebra as a
subalgebra) are classified \cite{sconf}. The highest possible dimension is six, which allows for the minimal conformal superagebra
(1,0) and the extended chiral conformal superalgebra (2,0).

In \cite{ISZ}, we construct and study the minimal superconformal $6D$ gauge theory. Unfortunately, it cannot be 
considered as a viable candidate for the TOE. It is not viable phenomenologically (the required 3-brane solutions are absent there)
and not internally self--consistent by two reasons:
 \begin{itemize}
  \item The Lagrangian involves among other things certain scalar fields of canonical dimension 2 with nontrivial {\it cubic}
potential.     It is not positive definite and the vacuum
is not classicaly stable. 
 \item Conformal symmetry is broken by anomaly in this theory. We have directly calculated the 1--loop $\beta$ function there and
obtained a non-vanishing answer.

\end{itemize}

Our hope  is that the extended $6D$ (2,0) theory is much better in this respect. 
We anticipate that $\beta$ function vanishes there to all orders (like it does in $4D$ ${\cal N} =4$ SYM) and that 
the quantum spectrum of this theory {\it has} a bottom. Maybe the mechanism by which the bottom (well-defined ground state)
appears is similar to the mechanism unravelled in the present paper. 
 Then  supersymmetry will distate as usual that the vacuum energy is zero if supersymmetry is not broken and positive
if it is. 

Unfortunately, no field theory with this symmetry group (the highest possible of all superconformal groups) is actually known now. 
The corresponding Lagrangian is not  constructed, and only indirect
results concerning scaling behavior of certain operators have been obtained so far using duality arguments 
\cite{M5}. Further studies of this very interesting
question are necessary.

\section*{Acknowledgements}
I am indebted to P. Mannheim, D. Robert, and  A. Vainshtein for  illuminating discussions and many valuable remarks.


\begin{thebibliography}{40}

\bibitem{PU} A. Pais and G.E. Uhlenbeck, Phys. Rev. {\bf 79}, 145 (1950).

\bibitem{duhi} A.V. Smilga, Nucl. Phys. {\bf B706}[PM], 598 (2005) \, [hep-th/0407031].

\bibitem{DM} P.D. Mannheim and A. Davidson, hep-th/0408104.

\bibitem{Ostr} M. Ostrogradsky, Mem. Ac. St. Petersburg {\bf VI 4}, 385 (1850).

\bibitem{Ham} T. Nakamura and S. Hamamoto, Progr. Theor. Phys. {\bf 95}, 469 (1996)\,  [hep-th/9511219].

\bibitem{fc} V. de Alfaro and T. Regge, {\it Potential Scattering}, Amsterdam (1965).\\
For a recent discussion, see A.E. Shabad, hep-th/0403177.

\bibitem{Bender} C.M. Bender, D.C. Brody and H.F. Jones, Am. J. Phys. {\bf 71}, 1095 (2003)\, [hep-th/0303005].

\bibitem{dragon}  A.V. Smilga, Phys. Atom. Nuclei {\bf 66} 2092 
(2003) [hep-th/0212033].  

 \bibitem{noncause} L. G\"odel, Rev. Mod. Phys. {\bf 21}, 447 (1949);
  A. Carlini and I.D. Novikov, { Int. J. Phys.}
{\bf D5}, 445  (1996) [gr-qc/9607063].

\bibitem{Isham}  See e.g. S. Hawking, { Comm. Math. Phys.} {\bf 87}, 
 395 (1982);  C.J. Isham, {\it Canonical Quantum Gravity and the Problem of Time}, 
gr-qc/9210011, published in the
Proceedings of GIFT Int. Seminar on Theor. Physics, Salamanca,
15-27 June, 1992.

\bibitem{sconf} For an instructive review, see S. Minwalla, Adv. Theor. Math. Phys.{\bf 2}, 781 (1998)  
[archive: hep-th/9712074].

\bibitem{ISZ} E.A. Ivanov, A.V. Smilga, and B.M. Zupnik, in preparation. 

\bibitem{M5}  P. Claus, R. Kallosh, and A. Van Proeyen, Nucl. Phys. {\bf B518}, 117 (1998)\, [hep-th/9711161];
O. Aharony, M. Berkooz, and N. Seiberg, Adv. Theor. Math. Phys. {\bf 2}, 119 (1998)\, 
[hep-th/9712117];
 F. Bastianelli, S. Frolov, and A.A. Tseytlin, JHEP {\bf 0002:013} (2000)\, [hep-th/0001041]; 
G. Arutyunov and E. Sokatchev, Nucl. Phys. {\bf B635} (2002) 332 \, [hep-th/0201145].



\end{thebibliography}
\end{document}